\documentclass[twocolumn,times]{aastex6}

\usepackage{ulem}
\usepackage{graphicx}
\usepackage{amsmath}
\usepackage{epsfig} 
\usepackage{enumitem}

\begin{document}

\title{Infrared emission from kilonovae: \\
the case of the nearby short hard burst GRB\,160821B}
\author{Mansi M. Kasliwal\altaffilmark{1}, Oleg Korobkin\altaffilmark{2}, Ryan M. Lau\altaffilmark{1}, Ryan Wollaeger\altaffilmark{2}, and Christopher L. Fryer\altaffilmark{2}}

\altaffiltext{1}{Division of Physics, Mathematics and Astronomy, California Institute of Technology, Pasadena, CA 91125, USA}
\altaffiltext{2}{Computational Methods Group (CCS-2), Los Alamos National Laboratory, P.O. Box 1663, Los Alamos, NM, 87545, USA}

\begin{abstract}
We present constraints on Ks-band emission from one of the nearest short hard gamma-ray bursts, GRB\,160821B, at z=0.16, at three epochs. We detect a reddened relativistic afterglow from the jetted emission in the first epoch but do not detect any excess kilonova emission in the second two epochs. We compare upper limits obtained with Keck I/MOSFIRE to multi-dimensional radiative transfer models of kilonovae, that employ composition-dependent nuclear heating and LTE opacities of heavy elements.  We discuss eight models that combine toroidal dynamical ejecta and two types of wind and one model with dynamical ejecta only. We also discuss simple, empirical scaling laws of predicted emission as a function of ejecta mass and ejecta velocity. Our limits for GRB\,160821B constrain the ejecta mass to be lower than 0.03 M$_{\odot}$ for velocities greater than 0.1\,c.  At the distance sensitivity range of advanced LIGO, similar ground-based observations would be sufficiently sensitive to the full range of predicted model emission including models with only dynamical ejecta. The color evolution of these models shows that I$-$K color spans 7--16\,mag, which suggests that even relatively shallow infrared searches for kilonovae could be as constraining as optical searches.
\end{abstract}


\section{Introduction}
Short Hard Bursts (SHBs) of $\gamma$-ray emission are purportedly neutron star mergers where the jet is conveniently pointed towards us (see \citealt{Berger14}
for a review). The same violent mergers are promising sources of gravitational wave emission in the advanced LIGO frequency band albeit independent of orientation 
and limited to the local Universe. 

The discovery of afterglows of SHBs has been much more challenging than that of long soft bursts due to their intrinsic faintness. In addition to an
afterglow, long bursts have been shown to be accompanied by a broad-line Type Ic supernova (e.g., \citealt{Galama98}). One may also expect ``kilonova" emission in SHBs from radioactive decay of heavy elements synthesized in these extreme environments (see \citealt{FM16} for a review). Kilonova emission could be red due to the opacities of heavy element lines \citep{kasen13,Barnes13}. Possible excess emission has also been reported for two nearby short bursts: excess optical emission was seen in GRB 080503  (albeit with a less secure redshift; \citealt{Perley09}) and infrared emission in GRB\,130603B \citep{Tanvir13,Fong14}. But the accompanying excess X-ray emission in both GRB\,080503 and GRB\,130603B cannot be easily explained by kilonova models. Additional claims of excess emission include GRB\,050709 \citep{Jin16} and GRB\,060614 \citep{Jin15}. Constraining upper limits on excess emission include GRB\,150101B \citep{Fong16}. 

The discovery of GRB\,160821B \citep{DiscoveryGCN} by the {\it Swift} satellite \citep{Gehrels04}, one of the nearest SHBs, presented another opportunity to look for any excess emission hinting at heavy element
radioactivity. An optical afterglow \citep{AfterglowGCN} and radio afterglow \citep{RadioGCN} were detected. A spectrum was obtained suggesting a very low redshift of 0.16  \citep{RedshiftGCN}.  Deep HST follow-up observations were also undertaken (\citealt{HSTGCN}, Troja et al. in prep). In this paper, we present a search for excess infrared emission at the Ks-band with Keck I/MOSFIRE. 

\section{Observations}
We obtained three epochs of deep imaging of GRB 160821B with the MOSFIRE \citep{MOSFIRE} instrument mounted on the Keck I telescope (Table~\ref{tab:data}) at
4.3 days, 7.5 days and 8.4 days after the $\gamma$-ray burst. Data were taken in correlated double sampling mode with an integration time of 4.4s and 7 co-adds in each exposure.
Multiple  well-dithered exposures were stacked on each night: specifically, 45 frames, 29 frames and 25 frames were stacked for the three epochs.  Data were reduced using standard procedures and calibrated relative to the 2MASS survey \citep{2MASS}. We marginally detect the afterglow in the first epoch (at 3$\sigma$) and report limiting magnitudes at the position of the afterglow of GRB\,160821B for the second two epochs. All data is summarized in Table~\ref{tab:data}.

\begin{deluxetable*}{llllll}
  \tabletypesize{\scriptsize}
  \tablecaption{Ks-band data on GRB\,160821B \label{tab:data}}
  \tablewidth{0pt}
  \tablehead{\colhead{MJD (Phase)} & \colhead{Instrument} & \colhead{Filter} & \colhead{Apparent mag (Vega)} & \colhead{Apparent mag (AB)} & \colhead{Absolute Mag (AB)}}
  \startdata
57626.234 (+4.3\,d) & Keck I/MOSFIRE & Ks & 22.19$^{+0.44}_{-0.31}$ & 24.04$^{+0.44}_{-0.31}$ & $-$15.4  \\
57629.402 (+7.5\,d) & Keck I/MOSFIRE & Ks  & $>$22.17 (3$\sigma$) & $>$24.02 (3$\sigma$) & $-$15.4 \\
57630.321 (+8.4\,d) & Keck I/MOSFIRE & Ks & $>$22.0 (3$\sigma$) & $>$23.85 (3$\sigma$) & $-$15.6 \\
 \enddata
\end{deluxetable*}

\section{Models}
\label{sec:models}

A wide variety of models exist in the literature that predict different kilonova signatures based on different opacity assumptions (e.g. \citealt{barnes16a,rosswog16a}). Here, we focus on a detailed comparison of our observations to only one family of models with the reddest opacity predictions (model details are described in \citealt{Wollaeger17}). This family of 9 models uses multi-dimensional radiative transfer simulations with a new treatment of multigroup opacity broadening, an approach described in \cite{fontes17a}. They are based on dynamical ejecta morphologies \citep{rosswog14a}, computed by long-term evolutions with radioactive heating source, following simulations of neutron star mergers in \citealt{rosswog13a}. Nucleosynthesis and radioactive heating are computed using nuclear network code {\tt WinNET}  \citep{winteler14,korobkin12} that is derived from the {\tt BasNet} network \citep{thielemann11}. Reaction rates for nucleosynthesis are taken from the \citealt{rauscher00a} compilation for the finite range droplet model (FRDM,  \citealt{moeller95a}), including density-dependent weak reaction rates \citep{arcones11a} and fission \citep{panov10a,panov05}. We calculate radioactive heating energy partitioning between different decay species and their thermalization, using empirical formulae derived in \citealt{barnes16a}. Radiative transfer simulations are performed using the open source code  {\tt SuperNu}\footnote{\url{https://bitbucket.org/drrossum/supernu/wiki/Home}} \citep{wollaeger14a}, which implements a 3D semi-implicit multigroup Monte Carlo solver.

The simplest model, dubbed {\tt SAd}, features spherically-symmetric analytic density distribution and contains dynamical ejecta only, with corresponding heating rates and opacities. This model is intended as the ``worst-case" scenario of a merger in which the brighter wind component is not present, or is obscured by the so-called ``lanthanide-curtain" \citep{kasen15a}, and also because this model is the least luminous due to its spherical shape with the lowest surface area. Thus, the {\tt SAd} model predicts the faintest kilonova signal.
  
The other eight models combine four different morphologies of dynamical ejecta (A-D from Table~1 in \cite{rosswog14a}) with two spherically-symmetric analytic models of wind, ``wind 1" and ``wind 2".  The wind carries two different nucleosynthetic compositions from representative tracers H5 and H1 in \citealt{perego14a}, respectively. These models are
  abbreviated as $\gamma A_1$, $\gamma B_1$, $\gamma C_1$, $\gamma D_1$, $\gamma A_2$, $\gamma B_2$, $\gamma C_2$ and  $\gamma D_2$.


Crude empirical scaling laws for the peak magnitudes (mp) in JHKs bands, depending on masses (m) and velocities (v), are as follows (for more details, see \citealt{Wollaeger17}):
\begin{equation}
mp_J = mp_{J0} - 0.93\,{\rm log} (m/m_0) - 1.61\,{\rm log} (v/v_0) 
\end{equation}
\begin{equation}
mp_H = mp_{H0} - 0.95\,{\rm log} (m/m_0) - 1.55\,{\rm log} (v/v_0)
\end{equation}
\begin{equation}
mp_K = mp_{K0} - 0.99\,{\rm log} (m/m_0) - 1.53\,{\rm log} (v/v_0)
\end{equation}

With these scaling formulas, starting with peak JHKs magnitudes {mp$_{J0}$, mp$_{H0}$, mp$_{K0}$} for a model with some mass m$_0$ and  some velocity v$_0$, we can estimate the peak magnitudes for a different mass and velocity.  Furthermore, we note that the time to maximum light also scales with ejecta mass and median velocity  \citep[c.f.][]{grossman14}: 
\begin{equation}
t_p/t_{p0}   = (m/m_0)^{0.32} (v/v_0)^{-0.60}
\end{equation}
These empirical fits are obtained using gray opacity models. The above models do not assume gray opacity and hence, we caution that these empirical fits are only an approximation.

%
%
%
%
%
%
%
%
%
%
%
%
%
%
%

\section{Discussion}

\begin{figure*}[!hbt]
\centering
\includegraphics[width=0.71\textwidth]{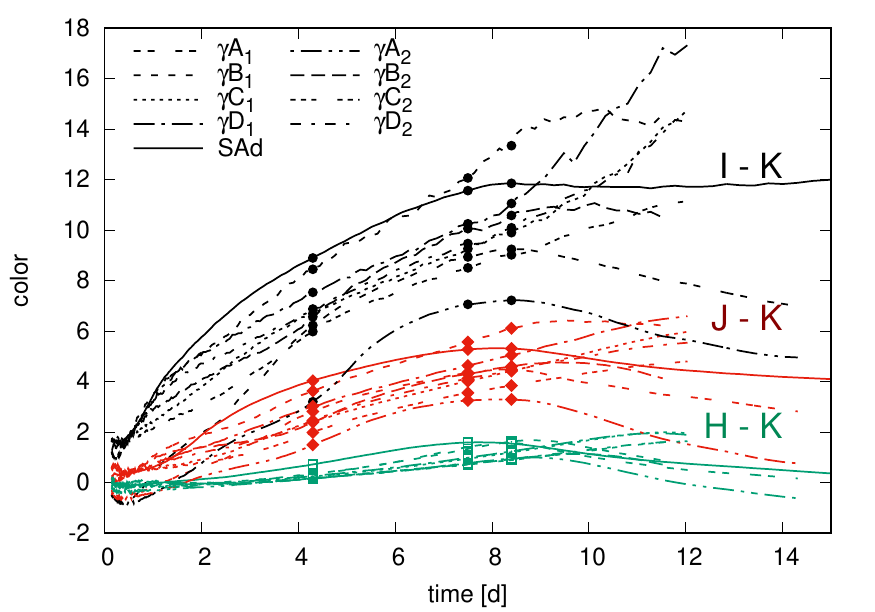}
\caption{I-K color (black circles), J-K color (red diamonds) and H-K color (green squares) evolution as a function of time since neutron star merger. Even a shallow infrared search for kilonovae would be more constraining than an optical search. H-band may be the most optimal ground-based infrared filter given the relatively lower sky brightness and small H-K colors.  
\label{fig:colors}}
\end{figure*}

\begin{figure*}[!hbt]
\centering
\includegraphics[width=0.72\textwidth]{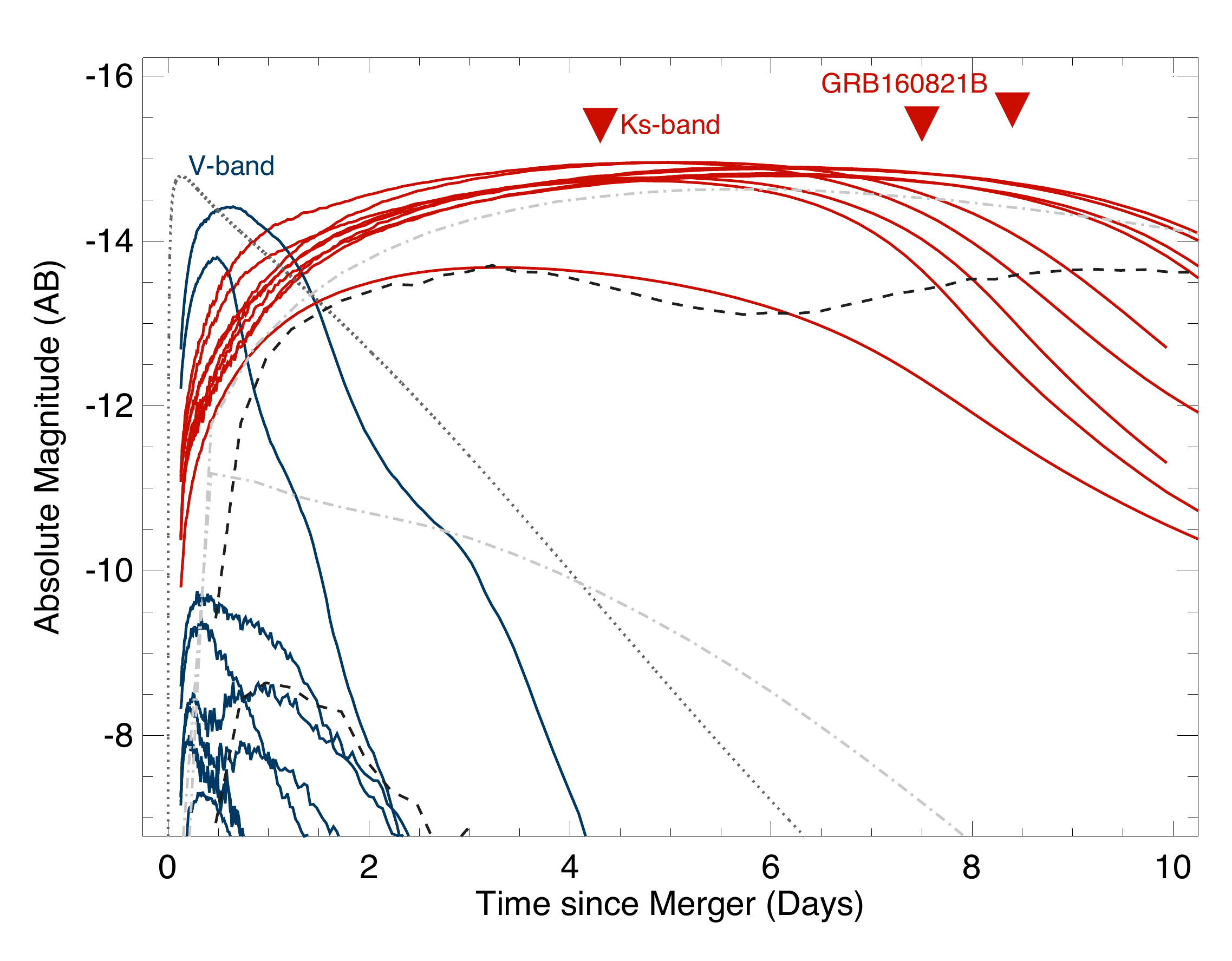}
\caption{Comparing the GRB160821B upper limits to model predictions in Ks-band (red solid lines) and V-band (blue solid lines) assuming ejecta mass of 0.01 M$_{\odot}$ and ejecta velocity of 0.1\,c. Also shown for comparison are other kilonova models in the literature with similar ejecta mass and ejecta velocity assumptions by \citealt{barnes16a} (black dashed line, Ks-band and V-band) and \citealt{rosswog16a} (light gray dash-dot line, Ks-band and V-band, ``ns12n14-dz31" model).  The bluest models are powered by beta decay of free neutrons (gray dotted line; \citealt{Metzger15}).   
\label{fig:lc}}
\end{figure*}

First,  we investigate whether our Ks-band detection of GRB\,160821B in the first epoch at 4.3\,d is consistent with afterglow emission from the jet. If we extrapolate V-mag of the afterglow reported in \citealt{HSTGCN} at 3.6\,d, and assume a power law decay based on fitting to afterglow photometry reported in \citealt{AfterglowGCN}, we find that the V$-$K color is approximately 1.9\,mag AB. This V$-$K color is somewhat redder than that expected from a typical SHB afterglow spectral index but it can possibly be explained by dust. Looking up the extensive compilation of afterglow data of short hard bursts by \citealt{Fong15},  we find that there is remarkably little Ks-band data that we can directly compare to (note that the latest phase of a near-IR afterglow detection in this compilation is only 1.5\,d).  The V$-$K color is inconsistent with all the kilonova models presented in this paper which predict colors redder than 6\,mag at this phase (Figure~\ref{fig:colors}). The V$-$K color is also not as red as models presented in \citealt{barnes16a} or \citealt{rosswog16a} (see Figure~\ref{fig:lc}). Furthermore, our Ks-band non-detections in the second two epochs at 7.5\,d and 8.7\,d are also consistent with the hypothesis that there was no detectable kilonova emission from this GRB (see Figure~\ref{fig:lc}). Additional contemporaneous multi-band multi-epoch photometry is necessary to securely disentangle whether or not there could be a contribution from both an afterglow and a kilonova at this 4.3\,d epoch. For example, if there was evidence to rule out a dust contribution, the excess emission could perhaps be explained by relatively bluer kilonova models with different assumptions on ejecta composition and ejecta opacity. 

\begin{figure*}[!hbt]
\centering
\includegraphics[width=0.72\textwidth]{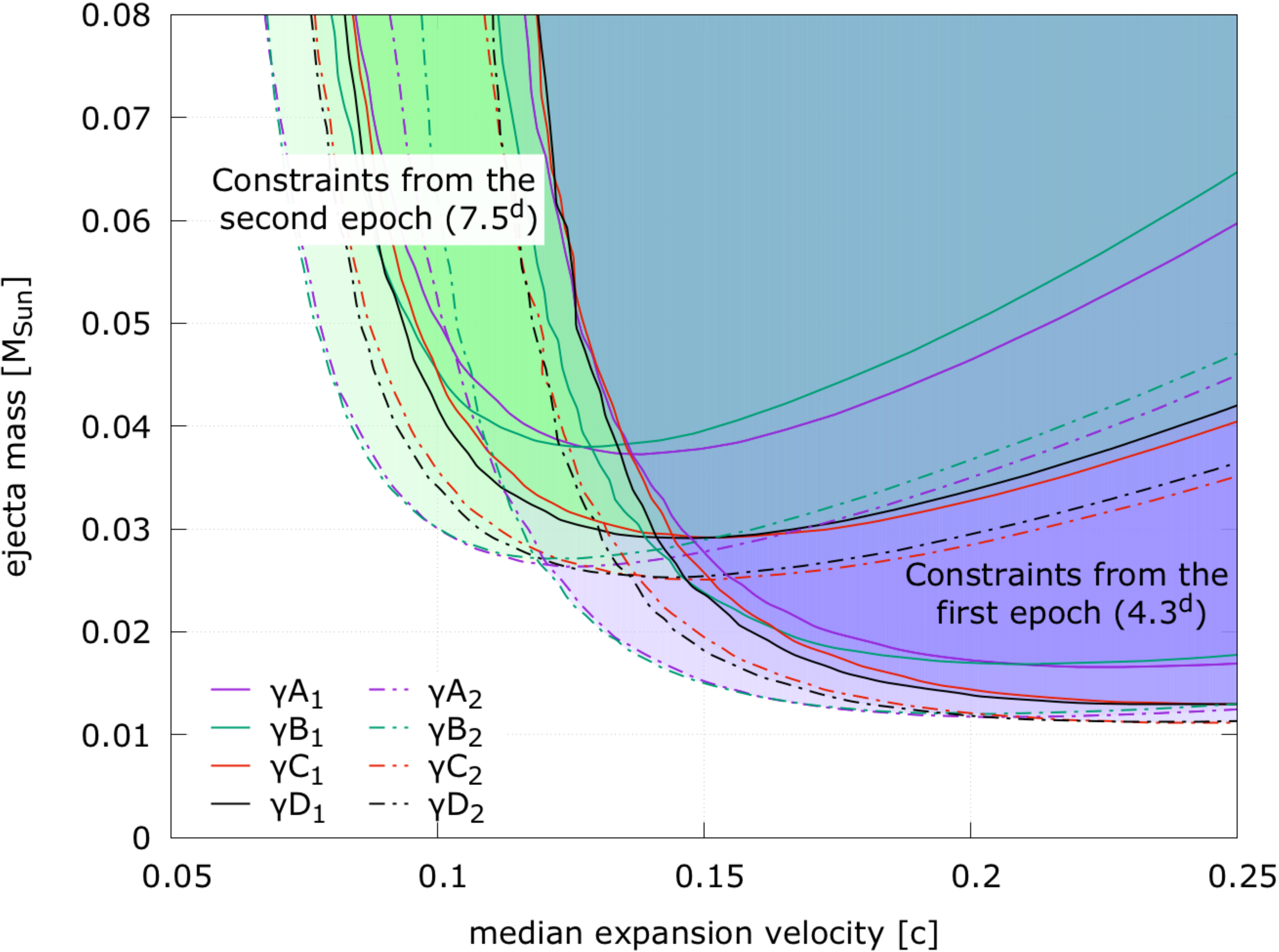}
\caption{The Ks-band upper limit for GRB\,160821B at 4.3\,d and 7.5\,d rules out the purple-shaded and green-shaded regions respectively with higher ejecta mass and higher ejecta velocity than denoted by the contours above. Each contour represents one of the eight models for each of two epochs. Assuming model predictions of dynamical ejecta in excess of 0.1\,c, our observations suggest ejecta masses lower than $0.03\ M_{\odot}$ for this event.
\label{fig:mvconstraints}}
\end{figure*}

Proceeding despite this caveat, we applied the empirical scaling laws discussed in \S~\ref{sec:models} to rescale the Ks-band light curves for each of our models to a different value of mass and velocity, for a range of masses and velocities. We then sampled the resulting redshifted light curves at the observed epochs and marked out regions with magnitudes excluded by upper limits at 4.3\,d and 7.5\,d. The result is shown in Figure~\ref{fig:mvconstraints}, where we plot excluded area for each model in the mass-velocity parameter space. The upper limit at epoch 8.7\,d does not provide additional constraints, so we do not show corresponding areas. Given that most simulations predict ejecta velocities higher than 0.1\,c \citep{hotokezaka13,rosswog13a}, we conclude that the mass of the dynamical ejecta of GRB\,160821B is less than 0.03 M$_{\odot}$. We note that this constraint is broadly consistent with other less red models in the literature \citep{barnes16a,rosswog16a} within a factor of few (see Figure~\ref{fig:lc}).  

\begin{figure*}[!hbt]
\centering
\includegraphics[width=0.5\textwidth]{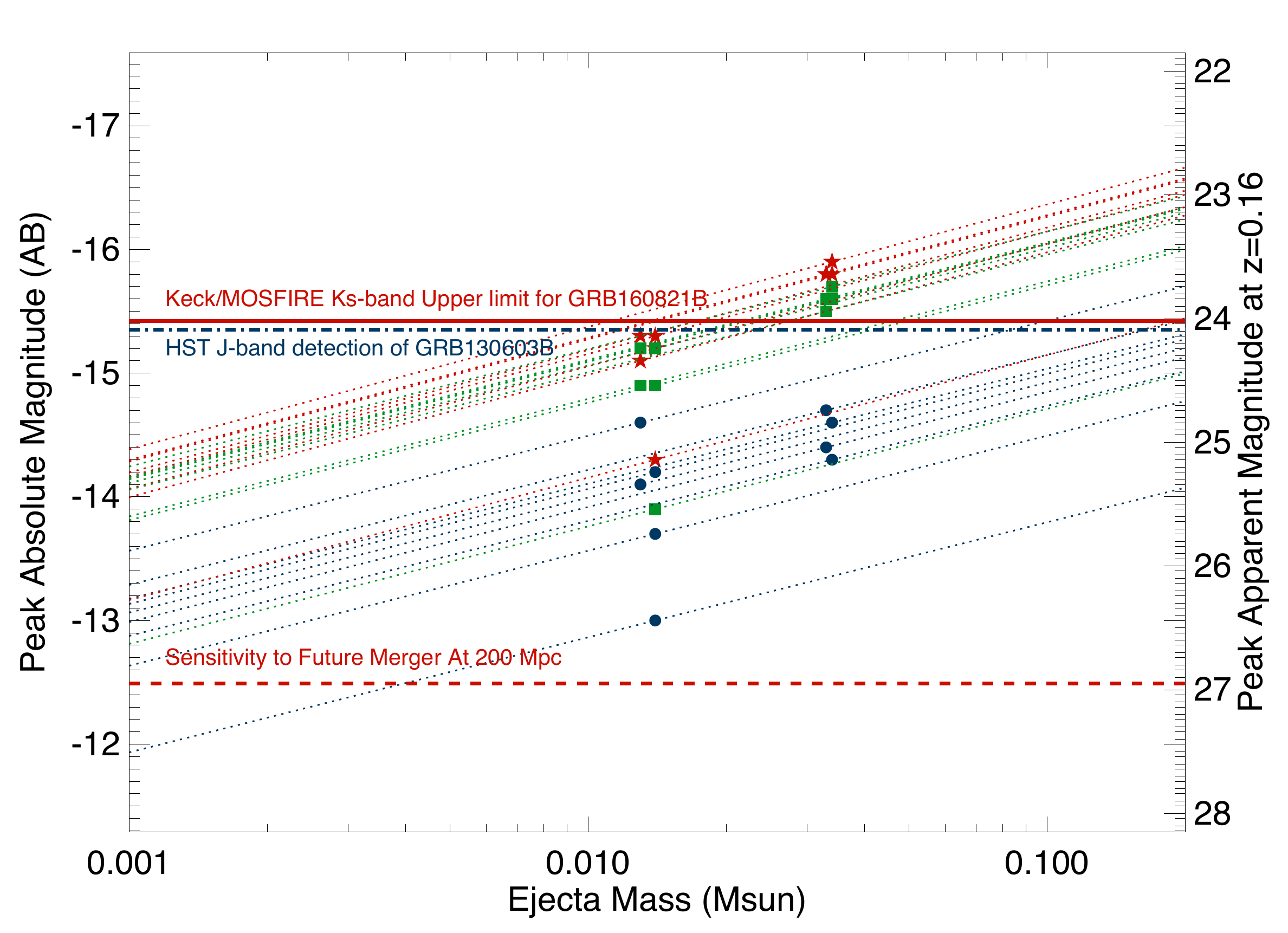}\includegraphics[width=0.5\textwidth]{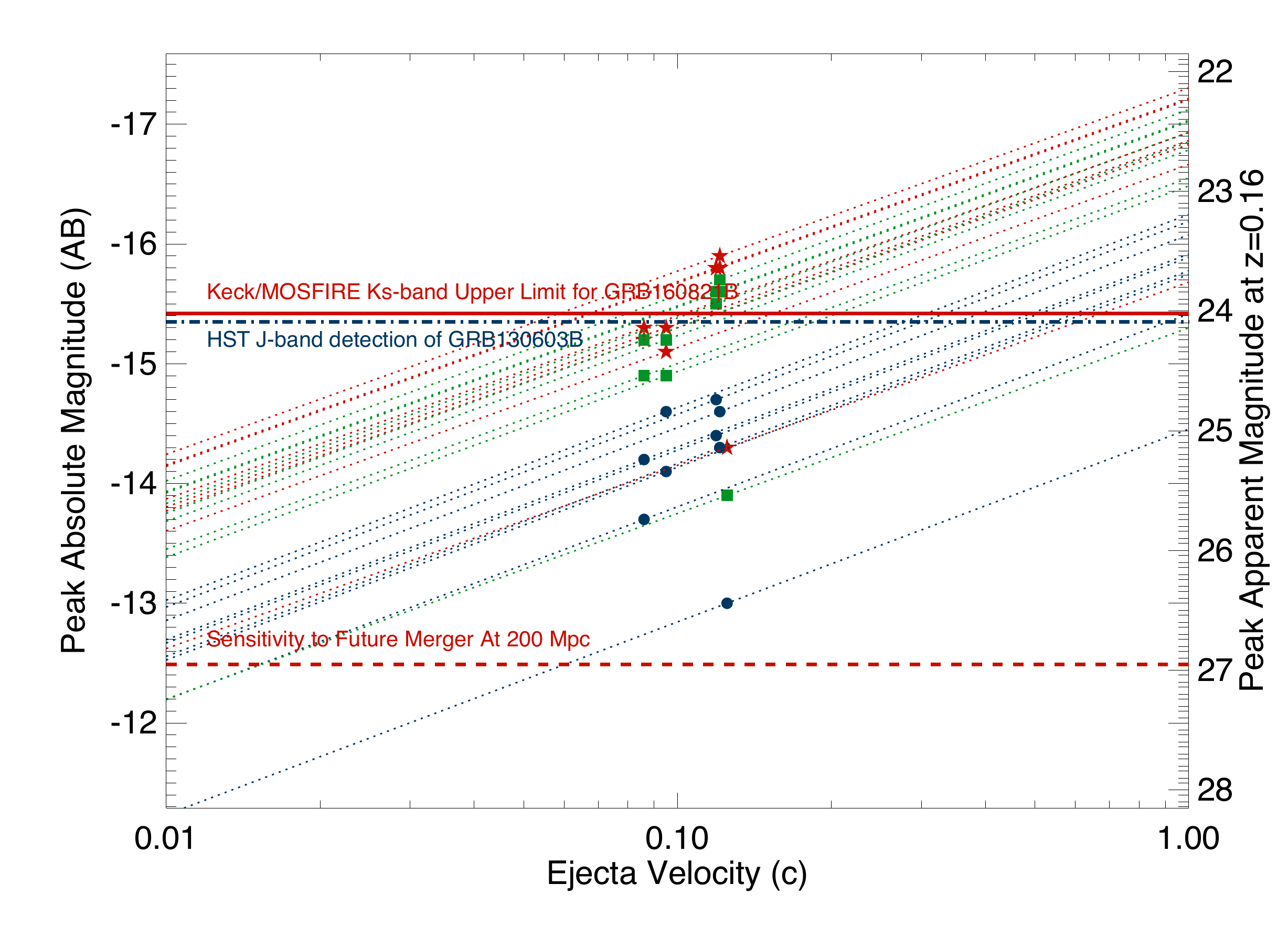}
\caption{Theoretical model predictions for J-band (blue circle), H-band (green square) and Ks-band (red star) as a function of ejecta mass (left panel) and ejecta velocity (right panel). Dotted lines show a rough empirical scaling with ejecta mass (left panel) and ejecta velocity (right panel). Our current Ks-band upper limit for GRB\,160821B constrains high ejecta mass and high ejecta velocity (solid red line). A future gravitational wave detection (dashed red line), at say 200 Mpc, would be so nearby that the same data would be able to test all 9 models covering the entire parameter space. Previously reported J-band detection of GRB\,130603B (dashed blue line) cannot be easily explained by the models studied in this paper. 
\label{fig:ejecta}}
\end{figure*}

To characterize the parameter space further of this family of models, we look at the predicted model emission as a function of ejecta mass and ejecta velocity using the empirical scaling laws described above  (see Figure~\ref{fig:ejecta}). Future gravitational wave detections of neutron star mergers will be relatively nearby due to the sensitivity of advanced gravitational wave interferometers being limited to approximately 200\,Mpc \citep{LVC16}. At this distance limit, similar ground-based Ks-band photometry would be extremely constraining for all models including the faintest SAd model with dynamical ejecta only (Figure~\ref{fig:ejecta}) for any assumption in ejecta mass or velocity spanning two orders of magnitude. 
For comparison on Figure~\ref{fig:ejecta}, we also show the J-band detection
of GRB130603B ($-$15.35 mag at 7\,d in comoving frame, \citealt{Tanvir13}).
The models presented here would suggest a very high ejecta
mass, $m_{ej}>0.08 M_{\odot}$ (or extreme velocities $>$0.4\,c 
which also shifts the peak to earlier time). However, one caveat here is that the nuclear mass model FRDM
tends to underestimate nuclear heating rates in comparison with other mass models \citep{dz95} and hence, produces dimmer kilonovae \citep{rosswog16a}.

Examining the extremely red color evolution further (Figure~\ref{fig:colors}), we find I$-$Ks colors peaking between 7--16\,mag. This suggests that even a relatively shallow infrared search for a kilonova would be competitive and complementary to constraints from optical searches (e.g., \citealt{Kasliwal16,Smartt16,Soares16}). The H$-$Ks color is relatively small, suggesting that either filter would work well.
Space-based observations, unhindered by night sky brightness, are deeper but are currently limited to H-band with narrow field-of-view cameras aboard the Hubble Space Telescope  (until the launch of the James Webb Space Telescope and Wide Field Infrared Survey Telescope). 

Infrared searches for kilonovae associated with coarsely localized gravitational wave triggers are currently inhibited by the astronomical cost of wide-field infrared detectors. The best effort currently is the 0.6 sq deg field-of-view camera on the 4m VISTA telescope \citep{Sutherland15}. For example, VISTA covered 8\% of the localization of GW\,150914 to a depth of J$<$20.7 (\citealt{Abbott16}, GCN\#18353). While the depth is constraining (corresponds to $-$14.3 at 100\,Mpc), the fractional area covered is too small. A future wide-field survey, say in the H-band or Ks-band to a similar depth, but covering a larger fraction of the error circle would be constraining in this context. We are exploring alternative semiconductors (\citealt{Sullivan14}, Simcoe et al. in prep) and/or creative optical design at a polar location \citep{Moore16} to break this cost barrier and/or blinding night-sky barrier to explore the dynamic infrared sky.  

%
%
\bigskip
MMK thanks M. Heida, F. Fuerst and E. S. Phinney for co-operating with the Target Of Opportunity interrupt observations at Keck I. We thank S. B. Cenko, J. Barnes, D. Kasen, R. Simcoe, D. A. Perley,  S. R. Kulkarni, W. Fong, N. Lloyd-Ronning and W. Even for valuable discussions. We thank our anonymous referee for helpful feedback. This work was supported by the GROWTH (Global Relay of Observatories Watching Transients Happen) project funded by the National Science Foundation Partnership in International Research Program under NSF PIRE grant number 1545949. Work at LANL was done under the auspices of the National Nuclear Security Administration of the U.S. Department of Energy at Los Alamos National Laboratory under Contract No. DE-AC52-06NA25396. All LANL calculations were performed on LANL Institutional Computing resources.
\facility{Keck:I (MOSFIRE)}

\end{document}